\documentclass[reqno,a4paper,12pt]{article}
\usepackage{amsmath}
\usepackage{amssymb}
\usepackage[dvips]{epsfig}
\usepackage{graphicx}

\usepackage{soul}

\begin{document}

\begin{center}
{\Large\bf LHC prospects in searches for neutral scalars in 
$pp\to\gamma\gamma+jet$: SM Higgs boson, radion, sgoldstino}
\\
\vspace{0.3cm}
S.~V.~Demidov$^{a,b,}$\footnote{{\bf e-mail}: demidov@ms2.inr.ac.ru}, 
D.~S.~Gorbunov$^{a,}$\footnote{{\bf e-mail}: gorby@ms2.inr.ac.ru}
\\

$^a${\small{\em 
Institute for Nuclear Research of the Russian Academy of Sciences, }}\\
{\small{\em
60th October Anniversary prospect 7a, Moscow 117312, Russia
}}

$^b${\small{\em 
Moscow Institute of Physics and Technology, }}\\
{\small{\em
Moscow region, Russia
}}
\\

\end{center}

\begin{abstract}
At hadron colliders the $\gamma\gamma+jet$ channel provides larger
signal-to-background ratio in comparison with inclusive $\gamma\gamma$
channel in hunting for scalars uncharged under the SM gauge group. At
NLO in QCD perturbation theory we evaluate selfconsistently the signal
significance for the SM Higgs boson production in $\gamma\gamma+jet$
channel at LHC. Three-body final state kinematics allows for refined
cuts. The adjustment of these cuts increases the signal significance
upto the level of inclusive channel.  Applying a justified simple
rescaling procedure to the results obtained for SM Higgs boson, we
estimate the LHC prospects in searches for radion and sgoldstino in
$\gamma\gamma+jet$ channel. We have found that this channel is
comparable with $\gamma\gamma$ channel in searches for new physics and
deserves further detailed investigations.

\end{abstract}

\section{Introduction}
With a lack of deeper understanding of strong interactions, search for
signal events in hadron collisions is somewhat ambiguous due to the
uncertainties in the prediction of the cross section of background
processes.  In case of large number of background events the
theoretical uncertainties in QCD predictions can result in swamping
the signal. Thus at hadron colliders among various channels of the
same signal significance the most favorable are those with large
signal-to-background ratio.

In this paper we study the LHC prospects in searches for Standard
Model (SM) Higgs boson, radion and sgoldstino in $\gamma\gamma+jet$
channel. In comparison with inclusive $\gamma\gamma$ channel, which
has almost the same signal significance, $\gamma\gamma+jet$ channel
exhibits larger signal-to-background ratio and, consequently, stronger
possibility to have control over QCD background. Though the signal
cross section is smaller in the channel with a high energy jet than in
the inclusive channel, rich 3-body kinematics in the final state
affords an opportunity to reduce the background significantly.

The $\gamma\gamma+jet$ channel has been extensively investigated as a
channel where SM Higgs boson with mass in the range $100-140$~GeV
should be discovered at LHC. At the leading order in perturbation
theory the signal significance was estimated in
Ref.~\cite{Abdullin:1998er}. The obtained results suggested that
$\gamma\gamma+jet$ channel is comparable with $\gamma\gamma$ channel
in searches for SM Higgs boson: the signal significances differ
slightly, while the signal-to-background ratio is several times larger
for $\gamma\gamma+jet$ channel.

This observation aroused the interest in phenomenology of this channel
and several thorough investigations have been performed.  In
particular, the NLO corrections to the cross sections of the dominant
signal subprocesses have been evaluated in
Ref.~\cite{deFlorian:1999zd} in the limit of the infinite mass of
$t$-quark, $M_t\to\infty$.  Similar to the inclusive channel, QCD
corrections double the Higgs production. The next natural step was
done in Ref.~\cite{deFlorian:1999tp,DelDuca:2003uz}, where cross
sections of main QCD background subprocesses were calculated at NLO
and their variations with cuts have been studied.  Finally, the photon
smooth isolation procedure~\cite{Frixione:1998jh}, which is required
to get rid of photons from fragmentation, was included in the analysis
of the NLO QCD background.

In this work we accumulate all relevant NLO results to calculate
selfconsistently the NLO signal significance for SM Higgs boson
production in $pp\to H(H\to\gamma\gamma)+jet$ channel. Along with
light $u,\;d,\;s$ quarks contribution we take into account also
antiquark and heavy quark contribution missed in paper
\cite{Abdullin:1998er} that raise the signal cross section by $6-7$\%
while background --- by about 25\%. We have found that for cuts
selected as in Ref.~\cite{Abdullin:1998er} the signal significance
remains almost intact with increasing order of perturbation
theory. Thus LO results for signal significance are stable with  
respect to QCD corrections. We estimate the uncertainties of our 
results for the signal significance to be not larger than 10\%, if
unknown higher order QCD corrections are disregarded. Let us stress
that our estimates are performed in a selfconsistent way with respect
to QCD perturbative calculus.   

Then we play with varying cuts and observe that their refinement
allows as much as 30\% rise in the signal significance, provided that
at LHC the very forward hadron calorimeters will operate properly.

The discovery of SM Higgs boson is a major goal of LHC. To
this end many channels have been scrutinized closely. One can exploit
these results to get for free accurate estimates of LHC sensitivity to
new physics which manifestation mimics SM Higgs boson
production. Indeed, any scalar uncharged under SM gauge group couples
to the SM fields exactly in the same way as SM Higgs boson. The only
distinguishing features are values of the corresponding coupling
constants, hence the production rate of the new scalar in the same
channel as SM Higgs boson production can be estimated by making use of
a simple rescaling procedure. Since the background is the same, this
yields the accurate estimate of the signal significance for the
production of this new scalar particle.

This method is applicable to many models, in particular, to any models
with extended Higgs sector. In this paper we consider two examples of
the appropriate models. The first example is provided by models with
warped extra spatial dimensions, where new scalar particle $\phi$,
radion, emerges in the low-energy spectrum. It is associated with
moduli, which vacuum expectation value $\Lambda_\phi$ suppresses
radion coupling to the SM fields. We estimate the LHC sensitivity to
$\Lambda_\phi$ in $\gamma\gamma+jet$ channel for the models with
radion mass of $100-140$~GeV. The second example is given by
supersymmetric extensions of the SM with low-energy spontaneous
supersymmetry breaking, where sgoldstinos gain masses of the order of
electroweak scale. Their couplings to the SM fields are determined by
the scale $\sqrt{F}$ of the supersymmetry breaking and corresponding
soft supersymmetry breaking terms. We estimate the LHC sensitivity to
$\sqrt{F}$ in $\gamma\gamma+jet$ channel for models with sgoldstino
mass of $100-300$~GeV and soft supersymmetry breaking terms within
$100-500$~GeV. Both examples suggest that discovery potential of
$\gamma\gamma+jet$ channel in searches for new physics is comparable
to prospects of the inclusive channel.  Note in passing that radion
production and sgoldstino production in inclusive channel have been
estimated for the first time~\cite{Giudice:2000av,Gorbunov:2002er} by
making use of the rescaling procedure similar to the one we apply in
this work.

The rest of the paper is organized as follows. In section II we
evaluate at NLO in perturbation theory the signal significance for SM
Higgs boson production at LHC in $pp\to\gamma\gamma+jet$
channel. There we study the dependence of the significance on
variations of the selected cuts and outline the optimal set of cuts.
Section III is devoted to estimates of LHC prospects in searches for
new physics in this channel. Namely, we consider multidimensional
models with radion of $100-140$~GeV and supersymmetric models with
sgoldstinos of $100-300$~GeV. Section IV contains discussion and
conclusions.

\section{Higgs boson}
We begin with studying LHC prospects in searches for SM Higgs boson in
$\gamma\gamma+jet$ channel. Since NLO $K$-factors for main signal and
background processes have been calculated
recently~\cite{deFlorian:1999zd,DelDuca:2003uz}, we embrace them to
improve the estimate~\cite{Abdullin:1998er} of the signal significance
by taking into account all relevant NLO corrections. In the next
sections we extend this result and estimate the LHC potential in
searches for radion and sgoldstino in this channel.

In calculations of the SM Higgs boson production at LHC we use the
CompHEP package~\cite{Pukhov:1999gg} with implemented $Hgg$, $Hggg$,
and $H\gamma\gamma$ effective point-like couplings. The coupling
constants entering these vertices have been obtained by matching the
corresponding partial widths, evaluated by means of CompHEP package,
with NLO results of HDECAY~\cite{Djouadi:1997yw}. This method is
justified because the analysis~\cite{deFlorian:1999zd} reveals that
NLO corrections to QCD subprocesses of $pp\to H(H\to\gamma\gamma)+jet$
can be reproduced by almost the same $K$-factor as for the inclusive
channel. Subprocesses with $WWH$ and $ZZH$ vertices are considered
only at tree level of perturbation theory. Although they give a
substantial contribution (about 20\%) to the signal cross section, we
do not expect any considerable deviation of total NLO results for
signal significance, since QCD corrections to SM Higgs boson
production via $W,Z$-fusion are rather modest,
$5-10$\%~\cite{Figy:2003nv}.

Evaluating the rates of both signal and background processes, we adopt
CTEQ6M approximation to NLO parton distribution functions, and for
main QCD subprocesses we set renormalization scale to
$Q^2=M_{\gamma\gamma}^2+(p_{T}^{jet})^2$, where $M_{\gamma\gamma}$ is
invariant mass of the photon pair and $p_{T}^{jet}$ is transverse
momentum of the hadronic jet. For the subprocesses with $W$- or
$Z$-bosons we set $Q^2=M_V^2$.

The NLO result for the background cross section was calculated in
Ref.~\cite{DelDuca:2003uz}, so to obtain the NLO approximation to the
signal significance we use almost the same set of cuts:
$p_{T}>40$~GeV, $|\eta|<2.5$ for both photons and jet with $\eta$
being pseudorapidity, $R_{\gamma\gamma}>0.4$, $R_{\gamma jet}>1.5$
(here $R_{ij}=\sqrt{\Delta\eta^2+\Delta\phi^2}$ is a separation
between two particles $i$ and $j$ in azimuth-angle--rapidity plane);
the isolation parameters are taken to be $R=1,\epsilon=0.5$, see
Ref.~\cite{Frixione:1998jh},~\cite{DelDuca:2003uz} for details.

The NLO results for the cross sections of the main signal subprocesses
and the background are presented in Table~\ref{main}~\footnote{
 Along with light $u,\;d,\;s$ quarks contribution we take into account also
antiquark and heavy quark contribution missed in paper
\cite{Abdullin:1998er} that raise the signal cross section by $6-7$\%
while background --- by about 25\%.}.
\begin{table}[htb]
\begin{center}
\begin{tabular}{|c|c|c|c|c|c|}
   \hline
   $M_{H}$, GeV   
   & 100 & 110 & 120 & 130 & 140 \\ \hline
   $gg\to g\gamma\gamma$, fb
   & 2.72 & 3.65 & 4.47 & 4.66 & 4.05 \\ \hline
   $qg\to q\gamma\gamma$, fb
   & 0.68 & 0.89 & 1.07 & 1.10 & 0.95   \\\hline
   $\bar{q}g\to\bar{q}\gamma\gamma$, fb
   & 0.38 & 0.49 & 0.59 & 0.60 & 0.52 \\   \hline
   $q\bar{q}\to g\gamma\gamma$, fb
   & 0.05 & 0.06 & 0.07 & 0.07 & 0.06 \\\hline
   W-,Z-contributions, fb
   & 1.23 & 1.67 & 1.86 & 1.91 & 1.68 \\\hline
   total signal cross section, $\sigma_{S}$, fb 
   & 5.06 & 6.76 & 8.06 & 8.34 & 7.25 \\\hline
   background cross section, $\sigma_{B}$, fb 
   &  53.2  & 55.6 &  56  & 57.3 & 55.6 \\\hline
   $N_{S}/N_{B}$
   & 0.10 & 0.12 & 0.14 & 0.15 & 0.13 \\\hline
\end{tabular}
\caption{\label{main} Summary of main signal and 
background cross sections. The background photons have been collected
within $M_{\gamma\gamma}\pm 1.4\cdot\sigma(M_{\gamma\gamma})$
interval, where $\sigma(M_{\gamma\gamma})$ is the mass resolution of
ATLAS detector~\cite{:1999fr}.}
\end{center}
\end{table}
The background events have been collected in a bin
$M_{\gamma\gamma}\pm 1.4\cdot\sigma(M_{\gamma\gamma})$, where
$\sigma(M_{\gamma\gamma})$ is the mass resolution of ATLAS
detector~\cite{:1999fr} (for CMS detector~\cite{unknown:1997kj} the
significance is designed to be higher by a factor of $1.4-1.5$).  Mass
range $100-115$~GeV is already experimentally excluded for SM Higgs
boson, but we will use these points in the next section to carry out
similar estimates for LHC sensitivity to new physics. It is worth to
note that in this estimates we do not take into account the efficiency
of photon and jet registrations in future LHC detectors.

\begin{figure}[htb]
\begin{center}
\includegraphics[width=0.8\columnwidth,height=0.40\columnwidth]{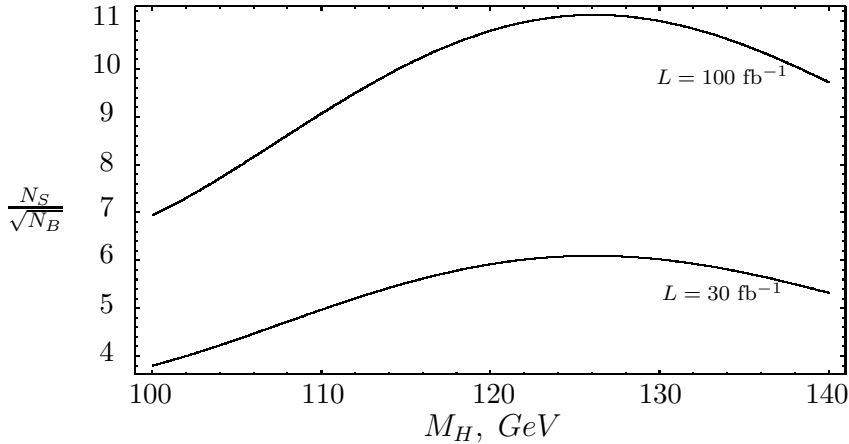}
\put(-313.00,160.00){\makebox(0,0)[cb]{{\small 11}}}
\put(-313.00,142.00){\makebox(0,0)[cb]{{\small 10}}}
\put(-313.00,124.00){\makebox(0,0)[cb]{{\small 9}}}
\put(-313.00,105.00){\makebox(0,0)[cb]{{\small 8}}}
\put(-313.00,88.00){\makebox(0,0)[cb]{{\small 7}}}
\put(-313.00,70.00){\makebox(0,0)[cb]{{\small 6}}}
\put(-313.00,51.00){\makebox(0,0)[cb]{{\small 5}}}
\put(-313.00,33.00){\makebox(0,0)[cb]{{\small 4}}}
\put(-296.50,17.00){\makebox(0,0)[cb]{{\small 100}}}
\put(-231.00,17.00){\makebox(0,0)[cb]{{\small 110}}}
\put(-166.50,17.00){\makebox(0,0)[cb]{{\small 120}}}
\put(-103.50,17.00){\makebox(0,0)[cb]{{\small 130}}}
\put(-40.50,17.00){\makebox(0,0)[cb]{{\small 140}}}
\put(-80.00,138.00){\makebox(0,0)[cb]{{\scriptsize $L=100~$fb$^{-1}$}}}
\put(-80.00,56.00){\makebox(0,0)[cb]{{\scriptsize $L=30~$fb$^{-1}$}}}
\put(-340.00,82.00){\makebox(0,0)[cb]{$\frac{N_{S}}{\sqrt{N_{B}}}$}}
\put(-167.00,2.00){\makebox(0,0)[cb]{$M_{H},\;GeV$}}
\caption{\label{higgssign}  
Signal significance $\frac{N_{S}}{\sqrt{N_{B}}}$ of the Higgs 
boson production in $pp\to\gamma\gamma+jet$ channel with integrated 
luminosities of $30$~fb$^{-1}$ and $100$~fb$^{-1}$.}
\end{center}
\end{figure}

The signal significance for ATLAS detector is plotted in
Fig.~\ref{higgssign} 
as a function of Higgs mass for two values of the integrated
luminosity: $30$~fb$^{-1}$ and $100$~fb$^{-1}$.  One can see, that in
the viable mass range, 115~GeV$<M_H<$140~GeV, it will be possible to
discover SM Higgs boson in $pp\to\gamma\gamma+jet$ channel even with
low luminosity of $30$~fb$^{-1}$ (the first year of LHC operating).

Comparing two channels, $pp\to H(H\to\gamma\gamma)+jet$ and $pp\to
H\to\gamma\gamma$, we find that ratio of the signal significances for
these channels (our results for the first channel and results,
presented in \cite{Bern:2002jx} for the second channel) is about
$0.8-0.9$, while the signal-to-background ratio is higher by a factor
of $2-3$ for the channel with a high energy jet. Hence, we confirm
that $\gamma\gamma+jet$ channel is competitive with the inclusive
channel in hunting for SM Higgs boson.

Let us discuss the accuracy of the presented results. The main
uncertainty is related to still unknown NNLO QCD corrections, or, in
our setup, to the strong dependence of the NLO results on the
renormalization scale. Indeed, it was shown in
Refs.~\cite{deFlorian:1999zd} and \cite{DelDuca:2003uz}, that for the
renormalization scale parameterized as $Q^2=x\cdot
(M_{\gamma\gamma}^2+(p_{T}^{jet})^2)$ the dominant QCD signal and
background cross sections decrease by about 50\% and $10-30$\%,
respectively, with $x$ changing from 0.5 to 2. With a lack of
understanding of the structure of QCD perturbation series, any
reliable quantitative estimate of this uncertainty seems improper.

Another source of uncertainty is associated with the local
approximation to $ggH$ coupling (valid in large $M_{t}$
limit). Namely, gluons and Higgs boson are assumed to be on
shell. Actually, this is not the case: one (two) gluon(s) entering LO
(NLO) diagrams can be off shell. However, for $M_{H}<M_{t}$ the local
approximation works well enough (with accuracy higher than a few
percent, see Refs.~\cite{DelDuca:2003ba}, \cite{Baur:1989cm}).

In our estimates we adopt the results for the background
processes~\cite{DelDuca:2003uz} obtained by making use of the photon
isolation procedure~\cite{Frixione:1998jh}. This procedure is aimed at
rejecting photons from fragmentation. For the inclusive $pp\to
H\to\gamma\gamma$ channel it was demonstrated in
Ref.~\cite{Bern:2002jx}, that with reasonable choice of the isolation
parameters the signal significance increases but mostly due to
reduction of the background.  In our calculations we do not apply this
procedure to single out the signal events and estimate the
corresponding corrections to be less than $5-10$\%.

Thus, without unknown QCD-corrections, we conclude that in total the
uncertainty of our results for the signal significance does not exceed
10\%. The obtained results and their accuracy ensure that SM Higgs
boson of mass within $115-140$~GeV will be definitely observed in
$pp\to\gamma\gamma+jet$ channel at LHC even with low integrated
luminosity of 30~fb$^{-1}$. 

Finally let us study the dependence of the signal significance on the 
cuts. We do not pretend on a completeness or 
on a high accuracy in this study, the
purpose is to catch the general tendency and outline the optimal set
of cuts. To this end we consider only dominant QCD-subprocesses, thus
neglecting $W$-boson and $Z$-boson contributions. Also we simplify the
evaluation as follows. The LO results for a new set of cuts are
obtained by making use of CompHEP package. The NLO corrections are
included as $K$-factors of the same values as for the old set of cuts.

To begin with, we include the additional cut on the energy of
scattering partons in their c.m.s., $\sqrt{\hat{s}}$, since this cut
enables one 
to reduce the background significantly~\cite{Abdullin:1998er}. The
results presented in Table~\ref{cut1} 
\begin{table}[htb]
\begin{center}
\begin{tabular}{|c|r@{.}l|r@{.}l|r@{.}l|r@{.}l|r@{.}l|}
  \hline
   $M_{H}$, GeV   & \multicolumn{2}{c|}{100} & \multicolumn{2}{c|}{110}
& \multicolumn{2}{c|}{120} & \multicolumn{2}{c|}{130} 
& \multicolumn{2}{c|}{140}\\ 
\cline{2-11} & 
\multicolumn{1}{c|}{$\frac{N_{S}}{\sqrt{N_{B}}}$} & 
\multicolumn{1}{c|}{$\frac{N_{S}}{N_{B}}$} & 
\multicolumn{1}{c|}{$\frac{N_{S}}{\sqrt{N_{B}}}$} & 
\multicolumn{1}{c|}{$\frac{N_{S}}{N_{B}}$} & 
\multicolumn{1}{c|}{$\frac{N_{S}}{\sqrt{N_{B}}}$} & 
\multicolumn{1}{c|}{$\frac{N_{S}}{N_{B}}$} & 
\multicolumn{1}{c|}{$\frac{N_{S}}{\sqrt{N_{B}}}$} & 
\multicolumn{1}{c|}{$\frac{N_{S}}{N_{B}}$} & 
\multicolumn{1}{c|}{$\frac{N_{S}}{\sqrt{N_{B}}}$} & 
\multicolumn{1}{c|}{$\frac{N_{S}}{N_{B}}$}
\\
\hline 
no cut on $\sqrt{\hat{s}}$ & 
\multicolumn{1}{c|}{5.3} & \multicolumn{1}{c|}{0.07} &
\multicolumn{1}{c|}{6.8} & \multicolumn{1}{c|}{0.09} &
\multicolumn{1}{c|}{8.3} & \multicolumn{1}{c|}{0.11} &
\multicolumn{1}{c|}{8.5} & \multicolumn{1}{c|}{0.11} &
\multicolumn{1}{c|}{7.5} & \multicolumn{1}{c|}{0.10} \\\hline
   $\sqrt{\hat{s}}>250$~GeV  & 
\multicolumn{1}{c|}{5.3} & \multicolumn{1}{c|}{0.12} &
\multicolumn{1}{c|}{6.7} & \multicolumn{1}{c|}{0.15} &
\multicolumn{1}{c|}{8.0} & \multicolumn{1}{c|}{0.15} &
\multicolumn{1}{c|}{8.1} & \multicolumn{1}{c|}{0.15} &
\multicolumn{1}{c|}{7.1} & \multicolumn{1}{c|}{0.12} \\\hline
   $\sqrt{\hat{s}}>300$~GeV & 
\multicolumn{1}{c|}{5.3} & \multicolumn{1}{c|}{0.16} &
\multicolumn{1}{c|}{6.7} & \multicolumn{1}{c|}{0.19} &
\multicolumn{1}{c|}{7.8} & \multicolumn{1}{c|}{0.20} &
\multicolumn{1}{c|}{7.8} & \multicolumn{1}{c|}{0.19} &
\multicolumn{1}{c|}{6.7} & \multicolumn{1}{c|}{0.15} \\\hline
   $\sqrt{\hat{s}}>350$~GeV & 
\multicolumn{1}{c|}{5.2} & \multicolumn{1}{c|}{0.20} &
\multicolumn{1}{c|}{6.5} & \multicolumn{1}{c|}{0.23} &
\multicolumn{1}{c|}{7.2} & \multicolumn{1}{c|}{0.24} &
\multicolumn{1}{c|}{7.6} & \multicolumn{1}{c|}{0.23} &
\multicolumn{1}{c|}{6.5} & \multicolumn{1}{c|}{0.18} \\\hline
\end{tabular}
\caption{\label{cut1} Dependence of the signal significance
$N_{S}/\sqrt{N_{B}}$ and signal-to-background ratio $N_{S}/N_{B}$ on
the additional cut on $\sqrt{\hat{s}}$ at the integrated luminosity of 
$100$~fb$^{-1}$.  The rest of the selection cuts is the same as in
Table~\ref{main}. The contributions from $W$- and $Z$-bosons have been
omitted.}
\end{center}
\end{table}
show that signal significance $N_{S}/\sqrt{N_{B}}$ always degrades
when cuts on $\sqrt{\hat{s}}$ are introduced.  On the other hand the
signal-to-background ratio increases from $0.09-0.11$ (without any cut
on $\sqrt{\hat{s}}$) to $0.2-0.25$ (for $\sqrt{\hat{s}}>350$~GeV).
Hence, additional cut on $\sqrt{\hat{s}}$ leads to considerable growth
of the signal-to-background ratio at a price of slight decrease in
signal significance.

Then we estimate  $N_{S}/\sqrt{N_{B}}$ for various cuts on
$p_{T}^{\gamma}$
(see Table~\ref{cut2}). 
\begin{table}[htb]
\begin{center}
\begin{tabular}{|c|r@{.}l|r@{.}l|r@{.}l|r@{.}l|r@{.}l|}
  \hline
   $M_{H}$, GeV   & \multicolumn{2}{c|}{100} & \multicolumn{2}{c|}{110}
& \multicolumn{2}{c|}{120} & \multicolumn{2}{c|}{130} 
& \multicolumn{2}{c|}{140}\\ 
\cline{2-11} & 
\multicolumn{1}{c|}{$\frac{N_{S}}{\sqrt{N_{B}}}$} & 
\multicolumn{1}{c|}{$\frac{N_{S}}{N_{B}}$} & 
\multicolumn{1}{c|}{$\frac{N_{S}}{\sqrt{N_{B}}}$} & 
\multicolumn{1}{c|}{$\frac{N_{S}}{N_{B}}$} & 
\multicolumn{1}{c|}{$\frac{N_{S}}{\sqrt{N_{B}}}$} & 
\multicolumn{1}{c|}{$\frac{N_{S}}{N_{B}}$} & 
\multicolumn{1}{c|}{$\frac{N_{S}}{\sqrt{N_{B}}}$} & 
\multicolumn{1}{c|}{$\frac{N_{S}}{N_{B}}$} & 
\multicolumn{1}{c|}{$\frac{N_{S}}{\sqrt{N_{B}}}$} & 
\multicolumn{1}{c|}{$\frac{N_{S}}{N_{B}}$}
\\
\hline 
$p_{T}^{\gamma}>40$~GeV  & 
\multicolumn{1}{c|}{5.3} & \multicolumn{1}{c|}{0.07} &
\multicolumn{1}{c|}{6.8} & \multicolumn{1}{c|}{0.09} &
\multicolumn{1}{c|}{8.3} & \multicolumn{1}{c|}{0.11} &
\multicolumn{1}{c|}{8.5} & \multicolumn{1}{c|}{0.11} &
\multicolumn{1}{c|}{7.5} & \multicolumn{1}{c|}{0.10} \\\hline
$p_{T}^{\gamma_{2}}>30$~GeV &
\multicolumn{1}{c|}{6.5} & \multicolumn{1}{c|}{0.07} &
\multicolumn{1}{c|}{8.2} & \multicolumn{1}{c|}{0.09} &
\multicolumn{1}{c|}{9.5} & \multicolumn{1}{c|}{0.11} &
\multicolumn{1}{c|}{9.4} & \multicolumn{1}{c|}{0.11} &
\multicolumn{1}{c|}{8.0} & \multicolumn{1}{c|}{0.10} \\\hline
 $p_{T}^{\gamma_{2}}>20$~GeV & 
\multicolumn{1}{c|}{6.9} & \multicolumn{1}{c|}{0.06} &
\multicolumn{1}{c|}{8.4} & \multicolumn{1}{c|}{0.08} &
\multicolumn{1}{c|}{9.6} & \multicolumn{1}{c|}{0.10} &
\multicolumn{1}{c|}{9.5} & \multicolumn{1}{c|}{0.10} &
\multicolumn{1}{c|}{8.0} & \multicolumn{1}{c|}{0.09} \\\hline

\end{tabular}
\caption{\label{cut2} Signal  
significance $N_{S}/\sqrt{N_{B}}$ signal-to-background ratio
$N_{S}/N_{B}$ at various cuts on $p_{T}^{\gamma}$ 
for one of photons with the integrated luminosity of $100$~fb$^{-1}$. 
The other selection cuts are the same as ones in the 
Table~\ref{main}. The contributions from $W$- and $Z$-bosons have been
omitted.
}
\end{center}
\end{table} 
The reason is that the standard set of cuts usually adopted for ATLAS
and CMS detectors in numerical simulations of the SM Higgs boson
production, $p_{T}^{\gamma_{1}}>40$~GeV, $p_{T}^{\gamma_{2}}>25$~GeV,
differs from our set. One can see, that adjustment of this cut yields
$10-15$\% increase in the signal significance (at a price of a
slight decrease in the signal-to-background ratio).

At last we vary cuts on $|\eta_{jet}|$, as motivated by 
the expected ability of hadronic calorimeter to cover the broad 
range of pseudorapidity, $|\eta|<4-5$. 
The results presented in Table~\ref{cut3}  
\begin{table}[htb]
\begin{center}
\begin{tabular}{|c|r@{.}l|r@{.}l|r@{.}l|r@{.}l|r@{.}l|}
  \hline
   $M_{H}$, GeV   & \multicolumn{2}{c|}{100} & \multicolumn{2}{c|}{110}
& \multicolumn{2}{c|}{120} & \multicolumn{2}{c|}{130} 
& \multicolumn{2}{c|}{140}\\ 
\cline{2-11} & 
\multicolumn{1}{c|}{$\frac{N_{S}}{\sqrt{N_{B}}}$} & 
\multicolumn{1}{c|}{$\frac{N_{S}}{N_{B}}$} & 
\multicolumn{1}{c|}{$\frac{N_{S}}{\sqrt{N_{B}}}$} & 
\multicolumn{1}{c|}{$\frac{N_{S}}{N_{B}}$} & 
\multicolumn{1}{c|}{$\frac{N_{S}}{\sqrt{N_{B}}}$} & 
\multicolumn{1}{c|}{$\frac{N_{S}}{N_{B}}$} & 
\multicolumn{1}{c|}{$\frac{N_{S}}{\sqrt{N_{B}}}$} & 
\multicolumn{1}{c|}{$\frac{N_{S}}{N_{B}}$} & 
\multicolumn{1}{c|}{$\frac{N_{S}}{\sqrt{N_{B}}}$} & 
\multicolumn{1}{c|}{$\frac{N_{S}}{N_{B}}$}
\\
\hline 
$|\eta_{jet}|<2.5$   & 
\multicolumn{1}{c|}{5.3} & \multicolumn{1}{c|}{0.07} &
\multicolumn{1}{c|}{6.8} & \multicolumn{1}{c|}{0.09} &
\multicolumn{1}{c|}{8.3} & \multicolumn{1}{c|}{0.11} &
\multicolumn{1}{c|}{8.5} & \multicolumn{1}{c|}{0.11} &
\multicolumn{1}{c|}{7.5} & \multicolumn{1}{c|}{0.10} \\\hline
$|\eta_{jet}|<4$ &
\multicolumn{1}{c|}{6.0} & \multicolumn{1}{c|}{0.08} &
\multicolumn{1}{c|}{7.9} & \multicolumn{1}{c|}{0.10} &
\multicolumn{1}{c|}{9.7} & \multicolumn{1}{c|}{0.12} &
\multicolumn{1}{c|}{10.1} & \multicolumn{1}{c|}{0.13} &
\multicolumn{1}{c|}{8.9} & \multicolumn{1}{c|}{0.11} \\\hline

\end{tabular}
\caption{\label{cut3}  
Signal significance $N_{S}/\sqrt{N_{B}}$ signal-to-background ratio
$N_{S}/N_{B}$ at various cuts on $\eta_{jet}$  
with the integrated luminosity of $100$~fb$^{-1}$. 
The other selection cuts are the same as in 
Table~\ref{main}. The contributions from $W$- and $Z$-bosons have been
omitted.
}
\end{center}
\end{table}
suggest that very forward calorimeter allows $10-15$\% raise in signal
 significance.

In total one can expect, that at least $10-30$\% increase in the
signal significance is anticipated at optimal choice of
cuts. Reverting to the comparison of $\gamma\gamma$ channel to
$\gamma\gamma+jet$ channel we conclude that the latter exhibits
practically the same signal significance as former and $2-2.5$ times
larger signal-to-background ratio. Consequently, $pp\to
H(H\to\gamma\gamma)+jet$ process can be treated even as a promising
alternative to the inclusive production.

\section{Other (pseudo)scalars}

Let us describe how to extend the analysis presented in the previous
section to the production of non-SM (pseudo)scalars. By making use of
the results obtained for SM Higgs boson we will estimate the
sensitivity of LHC in channel $pp\to\gamma\gamma+jet$ to the
multidimensional models with radions and supersymmetric models with
sgoldstinos.

The obvious procedure is a simple rescaling. Indeed, any scalar $X$
uncharged under SM gauge group interacts with SM massless gauge bosons
via nonrenormalizable couplings and the simplest among them are of the
same structure as for SM Higgs boson. The very values of the
corresponding coupling constants are the only difference. Hence, if
the gluon fusion mechanism dominates new-scalar particle production,
the signal cross section $\sigma_X$ for $pp\to
X(X\to\gamma\gamma)+jet$ process can be obtained by means of
rescaling 
\[
\sigma_X=\sigma_{H}\cdot \left(\frac{A_{Xgg}}{A_{Hgg}}\right)^2\cdot 
\frac{Br(X\to\gamma\gamma)}{Br(H\to\gamma\gamma)}\;, 
\]
where $A_{Hgg}$ and $A_{Xgg}$ are effective coupling constants
entering $Hgg$ and $Xgg$ vertices, respectively, and $\sigma_{H}$
contains only contributions from partonic diagrams with $ggH$ and
$gggH$ vertices to the Higgs boson production.  It is straightforward
to generalize this rescaling to the models with non-negligible $Xqq$
couplings. Certainly, $\sigma_X$ estimated in this way implies the
same set of cuts as $\sigma_{H}$.

Below we consider two different extensions of the SM with new
scalars. In both cases scalar production is dominated by gluon
fusion. For this special type of models the ratio of signal
significances of $pp\to X(X\to\gamma\gamma)$ and $pp\to
X(X\to\gamma\gamma)+jet$ channels is model independent: the only
relevant parameter governing this ratio is the scalar mass $M_X$. With
the standard set of cuts adopted in hunting for SM Higgs boson in the
inclusive channel~\cite{Bern:2002jx} and our set of cuts for
$\gamma\gamma+jet$ channel the ratio is about $0.6-0.7$ for
$M_X=100-140$~GeV.

Note in passing that, if subprocesses with exchange of $W$-boson or
$Z$-boson give a comparable contribution
to $X$ production, then: {\it (i)} rescaling procedure becomes 
rather involved, since both renormalizable and nonrenormalizable $XZZ$
and
$XWW$ couplings are generally allowed, while for SM Higgs boson only
renormalizable interactions exist, {\it (ii)} the universality of the
ratio of significances of two channels breaks down. However, in these
models hunting for scalar is more efficient in $WW^*$ and $ZZ^*$ decay
modes, that is beyond the scope of this paper.

\subsection{Radion}

In models with warped spatial extra dimensions (see, e.g., 
Ref.~\cite{Rubakov:2001kp} and
references therein), there is a module, radion, associated with
position of a brane. Stabilization of this
module~\cite{Goldberger:1999un} results in its coupling to the SM
fields
\[
\mathcal{L}_{\phi}=\frac{\phi}{\Lambda_{\phi}} T_{\mu}^{\mu}(SM)\;,
\]
where $\Lambda_\phi$ is a vacuum expectation value of the module and 
$T_{\mu}^{\mu}(SM)$ is the trace of SM energy-momentum tensor, which
consists of ordinary and anomaly contributions. The ordinary term is
\[
T_{\mu}^{\mu}(SM)^{ord}=\sum_{f} m_{f}\bar{f}
f-2m_{W}^2W^{+}_{\mu}W^{-\mu}-m_{Z}^2Z_{\mu}Z^{\mu}+...,
\]
where dots denote contributions of SM Higgs boson and higher order
terms. For gauge bosons there is also the anomaly contribution:
\[
T_{\mu}^{\mu}(SM)^{anom}=\sum_{all\; gauge\;
fields}\frac{\beta_{a}(g_{a})}{2g_{a}} F^{a}_{\mu\nu}F^{a\mu\nu}\;,
\]
where $\beta_a(g_a)$ are corresponding $\beta$-functions. 

The effective couplings of radion to gluons and photons 
are given by direct contribution from the trace anomaly and contribution
from the loop diagrams similar to the diagrams emerging in 
the case of SM Higgs
boson. As a result, $\phi g g$ coupling strongly dominates over $\phi
WW$ and $\phi ZZ$ couplings in comparison with the case of SM Higgs
boson: contribution of subprocesses with W- and Z- bosons is less than 
2\%. Hence our rescaling procedure is justified. 

\begin{figure}[htb]
\begin{center}
\includegraphics[width=0.95\columnwidth,height=0.50\columnwidth]{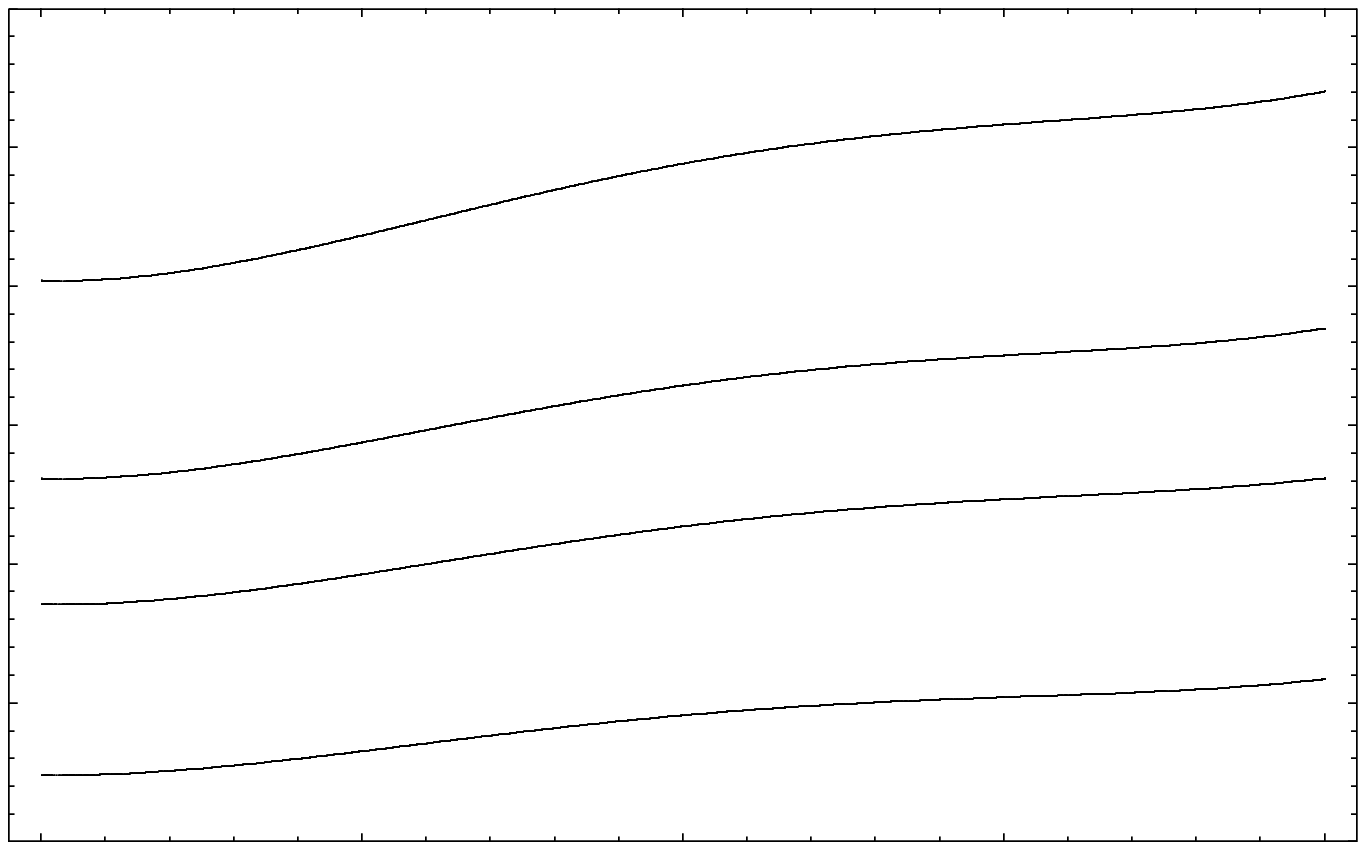}
\put(-342.00,81.00){\makebox(0,0)[cb]{{\small 2}}}
\put(-342.00,102.00){\makebox(0,0)[cb]{{\small 2.5}}}
\put(-342.00,126.00){\makebox(0,0)[cb]{{\small 3}}}
\put(-342.00,148.00){\makebox(0,0)[cb]{{\small 3.5}}}
\put(-342.00,171.00){\makebox(0,0)[cb]{{\small 4}}}
\put(-342.00,194.00){\makebox(0,0)[cb]{{\small 4.5}}}
\put(-324.00,49.00){\makebox(0,0)[cb]{{\small 100}}}
\put(-261.00,49.00){\makebox(0,0)[cb]{{\small 110}}}
\put(-199.50,49.00){\makebox(0,0)[cb]{{\small 120}}}
\put(-137.50,49.00){\makebox(0,0)[cb]{{\small 130}}}
\put(-74.00,49.00){\makebox(0,0)[cb]{{\small 140}}}
\put(-380.00,124.00){\makebox(0,0)[cb]{$\Lambda_{\phi},\;TeV$}}
\put(-200.00,34.00){\makebox(0,0)[cb]{$M_{\phi},\;GeV$}}
\put(-200.00,176.00){\makebox(0,0)[cb]{{\scriptsize $L=500~$fb$^{-1}$}}}
\put(-200.00,140.00){\makebox(0,0)[cb]{{\scriptsize $L=200~$fb$^{-1}$}}}
\put(-200.00,116.00){\makebox(0,0)[cb]{{\scriptsize $L=100~$fb$^{-1}$}}}
\put(-200.00,85.00){\makebox(0,0)[cb]{{\scriptsize $L=30~$fb$^{-1}$}}}
\caption{\label{radion} Signal significance (5$\sigma$-level) 
for the radion production 
in $pp\to\gamma\gamma+jet$ reaction with integrated 
luminosity $30$~fb$^{-1}$, $100$~fb$^{-1}$ and $200$~fb$^{-1}$.}
\end{center}
\end{figure}

In models with radion there are only two free parameters~\footnote{ In
a number of models higgs-radion mixing can arise, but in this paper we
ignore this possibility.}: radion mass $m_\phi$ and $\Lambda_\phi$
(current experimental bounds~\cite{Hagiwara:fs} are $M_{\phi}>120$~GeV
at $\Lambda_{\phi}=1$~TeV).  So we investigate the LHC sensitivity in
$pp\to\gamma\gamma+jet$ channel to the scale $\Lambda_{\phi}$ in
models with radion mass $m_\phi=100-140$~GeV by rescaling results
obtained for SM Higgs boson, as explained above.  The results for
the models with radion are presented in Fig.~\ref{radion}
for a set of integrated luminosities. Any models
with parameters in 
the region below plotted lines will be discovered at
LHC (ATLAS detector only) at the confidence level better than
5$\sigma$ if corresponding integrated luminosities are collected. 
One can conclude that it is possible to discover radion with masses of
$100-140$~GeV in $pp\to\gamma\gamma+jet$ channel, if stabilization scale
is not higher than $4$~TeV. 
 
\subsection{Sgoldstino}

The models with spontaneous supersymmetry breaking contain goldstino
supermultiplet, which includes scalar particles --- sgoldstinos ---
superpartners of goldstino. In a variety of models (see, e.g.,
Refs.~\cite{Ellis:1984kd}, \cite{Giudice:1998bp}) these particles are
relatively light and could be produced at LHC. The part of sgoldstino
interaction terms relevant~\footnote{Here we consider only scalar
sgoldstino. It is straightforward to extend the results presented in
this section to the production of pseudoscalar sgoldstino as well.
The sensitivity of the channel under discussion to coupling constants
of pseudoscalar sgoldstino coincides with the sensitivity to the
coupling constants of the scalar sgoldstino of the same mass.}  to the
study of sgoldstino production in $pp\to\gamma\gamma+jet$ channel
reads~\cite{Gorbunov:2001pd}
\begin{align}
\mathcal{L}_{S}=-\sum_{all\; gauge\; fields}
\frac{M_{\alpha}}{2\sqrt{2}F}S\cdot F^{\alpha}_{a \mu\nu}F^{\alpha
\mu\nu}_{a} 
\nonumber
- 
\frac{\mathcal{A}^{L}_{ab}}{\sqrt{2}F}y^{L}_{ab}\cdot
S(\epsilon_{ij}l^{j}_{a}e^{c}_{b}h_{D}^{i}+h.c.)
\\\label{eq1}
-
\frac{\mathcal{A}^{D}_{ab}}{\sqrt{2}F}y^{D}_{ab}\cdot
S(\epsilon_{ij}q^{j}_{a}d^{c}_{b}h_{D}^{i}+h.c.)-
\frac{\mathcal{A}^{U}_{ab}}{\sqrt{2}F}y^{U}_{ab}\cdot
S(\epsilon_{ij}q^{j}_{a}u^{c}_{b}h_{U}^{i}+h.c.),
\end{align}
where $M_{\alpha}$ are gaugino masses, $A_{ab}y_{ab}$ are soft
trilinear coupling constants, $\sqrt{F}$ is the scale of supersymmetry
breaking, and $\epsilon_{ij}$ is $2\times2$ antisymmetric matrix fixed
as $\epsilon_{12}=1$. The current experimental
bounds~\cite{Abreu:2000ij} are
$\sqrt{F}>500-200$~GeV at $M_{S}>10-150$~GeV with MSSM soft mass terms
$M_{soft}$ being of order 100~GeV. 

Since $Sgg$, $SWW$ and $SZZ$ coupling constants are of the same order,
sgoldstino production is saturated by gluon fusion (diagrams with
$W,Z$-bosons involve additional weak vertices and corresponding
contributions are less than a few percent). Yukawa-type coupling
is important for sgoldstino interactions with $t$-quarks, and also for
sgoldstino interactions with $b$-quarks, if $\tan\beta$ is
sufficiently large.  However, sea $b$-quarks carry only a small
fraction of proton momentum, and their contributions to sgoldstino
production can be neglected. Hence one can apply directly the
rescaling procedure to estimate LHC prospects in searches for
sgoldstino in $pp\to\gamma\gamma+jet$ channel.

For the models with sgoldstino we present the estimates of the
LHC sensitivity to the scale of supersymmetry breaking $\sqrt{F}$
for the same sets of MSSM soft parameters as ones 
considered in~\cite{Gorbunov:2002er}, see Table~\ref{parameters}. 
\begin{table}[htb]
\begin{center}
\begin{tabular}{|c|c|c|c|c|}
\hline
Model & $M_{1}$ & $M_{2}$ & $M_{3}$ &  $A$ \\\hline
  I   & 100 GeV & 300 GeV & 500 GeV & 300 GeV \\\hline
 II   & 300 GeV & 300 GeV & 300 GeV & 300 GeV \\\hline
\end{tabular}
\caption{\label{parameters} The values of MSSM soft terms for two
supersymmetric models.}
\end{center}
\end{table}

For SM Higgs boson $\gamma\gamma$ decay mode is out of interest for
$M_H>140$~GeV, since Higgs width starts to grow rapidly with its
mass, thereby diminishing $Br(H\to\gamma\gamma)$. The similar
situation takes place in models with radion. Quite the contrary, 
sgoldstino width is generally saturated by decay into gluons, so
two-photon
branching ratio remains intact for $M_S\simeq100-300$~GeV, and
$pp\to\gamma\gamma+jet$ channel (as well as inclusive one) may be
employed for searches for sgoldstino not only in the mass range
relevant for SM Higgs boson, $115-140$~GeV, but 
also in a wider region. 
While in models with $M_S\lesssim140$~GeV one can apply the rescaling
procedure to obtain the LHC sensitivity to sgoldstino couplings, the
opposite case, $M_S\gtrsim140$~GeV, requires a special study. Indeed,
NLO background and $\gamma\gamma$-invariant mass resolution as well as
photon isolation procedure have not been thoroughly analyzed for this
mass interval. To estimate the LHC prospects in searches for
sgoldstino with $M_S\gtrsim140$~GeV, we adopt the photon energy
resolution of ATLAS electromagnetic calorimeter~\cite{:1999fr} for
$\gamma\gamma$-invariant mass resolution. Both cross section of
sgoldstino production and background cross section are 
calculated at LO by making use of CompHEP-sgoldstino
package~\cite{Gorbunov:2001pd}. Finally, these results are corrected
by NLO K-factors, which are treated as constants in the whole mass 
region, $100-300$~GeV. 

\begin{figure}[htb]
\includegraphics[width=1.0\columnwidth,height=0.45\columnwidth]{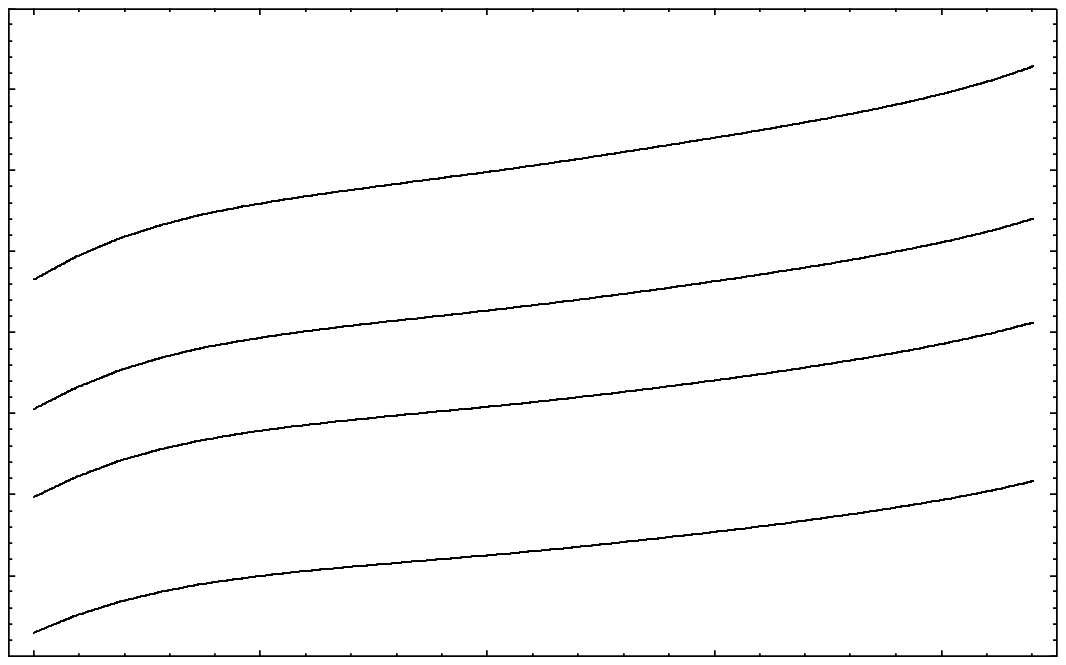}
\put(-202.00,0.00){\makebox(0,0)[cb]{{\small $M_{S},\;GeV$}}}
\put(-370.00,99.00){\makebox(0,0)[cb]{{\small $\sqrt{F},\;TeV$}}}
\put(-182.00,144.00){\makebox(0,0)[cb]{{\scriptsize $L=500~$fb$^{-1}$}}}
\put(-182.00,112.00){\makebox(0,0)[cb]{{\scriptsize $L=200~$fb$^{-1}$}}}
\put(-182.00,91.00){\makebox(0,0)[cb]{{\scriptsize $L=100~$fb$^{-1}$}}}
\put(-182.00,59.00){\makebox(0,0)[cb]{{\scriptsize $L=30~$fb$^{-1}$}}}
\put(-202.00,17.00){\makebox(0,0)[cb]{{\small 200}}}
\put(-145.00,17.00){\makebox(0,0)[cb]{{\small 250}}}
\put(-88.00,17.00){\makebox(0,0)[cb]{{\small 300}}}
\put(-258.00,17.00){\makebox(0,0)[cb]{{\small 150}}}
\put(-315.00,17.00){\makebox(0,0)[cb]{{\small 100}}}
\put(-333.00,45.00){\makebox(0,0)[cb]{{\small 5.5}}}
\put(-333.00,64.00){\makebox(0,0)[cb]{{\small 6}}}
\put(-333.00,80.70){\makebox(0,0)[cb]{{\small 6.5}}}
\put(-333.00,100.00){\makebox(0,0)[cb]{{\small 7}}}
\put(-333.00,117.50){\makebox(0,0)[cb]{{\small 7.5}}}
\put(-333.00,135.60){\makebox(0,0)[cb]{{\small 8}}}
\put(-333.00,153.50){\makebox(0,0)[cb]{{\small 8.5}}}
\put(-333.00,171.50){\makebox(0,0)[cb]{{\small 9}}}
\caption{\label{sgold1} $5\sigma$-level discovery contours for 
sgoldstino production (Model I) in $pp\to\gamma\gamma+jet$ channel with
various 
values of integrated luminosity; the set of soft supersymmetry
breaking parameters is listed in Table~\ref{parameters}.}
\end{figure}

\begin{figure}[htb]
\includegraphics[width=1.0\columnwidth,height=0.45\columnwidth]{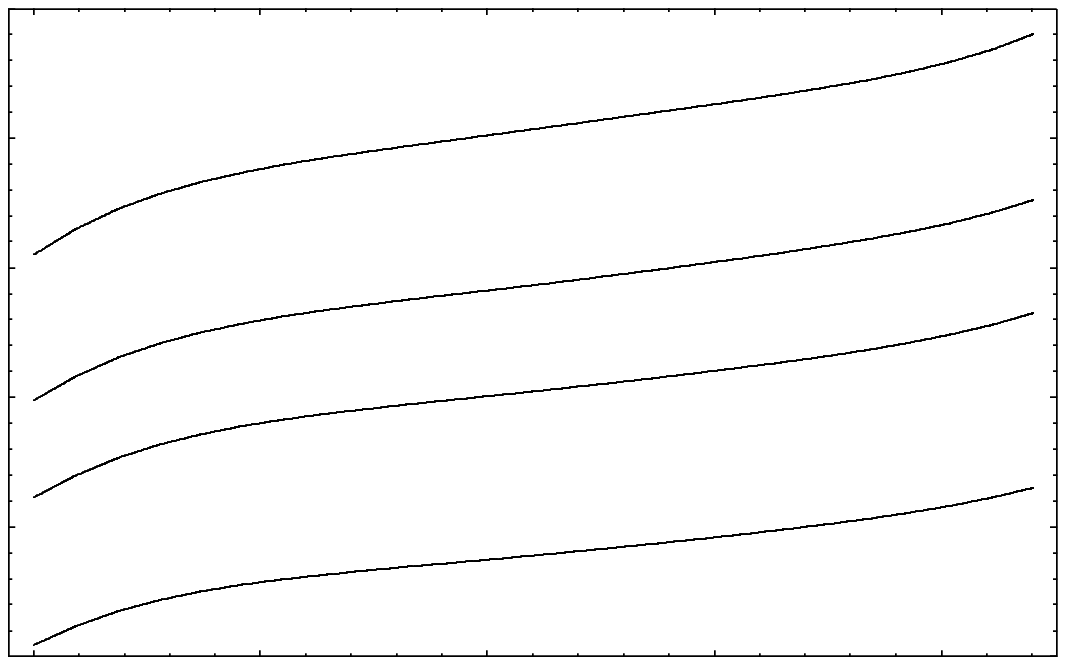}
\put(-214.00,0.00){\makebox(0,0)[cb]{{\small $M_{S},\;GeV$}}}
\put(-380.00,100.00){\makebox(0,0)[cb]{{\small $\sqrt{F},\;TeV$}}}
\put(-182.00,153.00){\makebox(0,0)[cb]{{\scriptsize $L=500~$fb$^{-1}$}}}
\put(-182.00,118.00){\makebox(0,0)[cb]{{\scriptsize $L=200~$fb$^{-1}$}}}
\put(-182.00,94.00){\makebox(0,0)[cb]{{\scriptsize $L=100~$fb$^{-1}$}}}
\put(-182.00,58.00){\makebox(0,0)[cb]{{\scriptsize $L=30~$fb$^{-1}$}}}
\put(-214.00,17.00){\makebox(0,0)[cb]{{\small 200}}}
\put(-157.00,17.00){\makebox(0,0)[cb]{{\small 250}}}
\put(-100.00,17.00){\makebox(0,0)[cb]{{\small 300}}}
\put(-270.00,17.00){\makebox(0,0)[cb]{{\small 150}}}
\put(-328.00,17.00){\makebox(0,0)[cb]{{\small 100}}}
\put(-343.00,170.00){\makebox(0,0)[cb]{{\small 12}}}
\put(-343.00,142.00){\makebox(0,0)[cb]{{\small 11}}}
\put(-343.00,113.00){\makebox(0,0)[cb]{{\small 10}}}
\put(-343.00,84.50){\makebox(0,0)[cb]{{\small 9}}}
\put(-343.00,56.50){\makebox(0,0)[cb]{{\small 8}}}
\caption{\label{sgold2} $5\sigma$-level discovery contours for
sgoldstino production (Model II) in $pp\to\gamma\gamma+jet$ channel
with various values of integrated luminosity; the set of soft
supersymmetry breaking parameters is listed in
Table~\ref{parameters}.}
\end{figure}

The results for the two models are presented in Figs.~\ref{sgold1} and
\ref{sgold2}.
The solid lines indicate the scales of supersymmetry
breaking $\sqrt{F}$ which will be tested at 5$\sigma$-level in
searches for sgoldstino in $pp\to\gamma\gamma+jet$ channel at LHC
(ATLAS detector only) with various integrated luminosities. One can
see, that the scale of supersymmetry breaking $\sqrt{F}$ will be probed
up to $8-12$~TeV depending on the pattern of MSSM soft terms and 
sgoldstino mass.

\section{Discussion and Conclusions}

In this work we have explored the capability of LHC in searches
for SM Higgs boson, sgoldstino and radion in $pp\to\gamma\gamma+jet$
channel.  The NLO effects in both the signal and the background cross
sections were taken into account in the selfconsistent local
approximation for $ggH$ coupling.

We have confirmed that SM Higgs boson of $115-140$~GeV will be
discovered at LHC in this channel even with low integrated luminosity
of $30$~fb$^{-1}$.  Comparing the potentials of
$pp\to\gamma\gamma+jet$ and $pp\to\gamma\gamma$ channels, we have
found that the ratio of signal significances of these channels is
about $0.8-0.9$, while the signal-to-background ratio is larger by a
factor of $2-3$ for the channel with a high energy jet. The
uncertainties of the obtained results are expected to be less than
10\%, neglecting unknown higher order QCD corrections.  It was
shown that tuning of cuts on $p_{T}^{\gamma}$ and $\eta_{jet}$ in this
channel could yield $10-30$\% enhancement of the signal significance.
This suggests that $\gamma\gamma+jet$ channel is highly competitive with the
inclusive one. Moreover, with account of larger signal-to-background
ratio $\gamma\gamma+jet$ channel seems even more favorable. The
definite answer requires further detailed investigations. In
particular, one has to take into consideration that $ggH$ effective
coupling is nonlocal in this process, since at least one of the gluons
is off shell. Likewise we did not take into consideration the
registration efficiency of the future LHC detectors.

Starting from the results for SM Higgs boson and adopting the
rescaling procedure we have estimated the LHC prospects in searches
for new physics in the channel with a high energy jet. For models with
warped extra dimensions we have observed that radion with mass of
$100-140$~GeV could be discovered in $\gamma\gamma+jet$ channel, if
stabilization scale $\Lambda_\phi\lesssim4$~TeV. In models with
low-energy supersymmetry and sgoldstino masses of $100-300$~GeV the
scale of supersymmetry breaking $\sqrt{F}$ could be probed in
$\gamma\gamma+jet$ channel up to about $8-12$~TeV (depending on the
MSSM parameters of soft supersymmetry breaking). These results ensure
that $pp\to\gamma\gamma+jet$ channel is very promising in searches for
new physics.

The proposed rescaling procedure can be applied for estimates of LHC
sensitivity to various extensions of the SM with new (pseudo)scalars
uncharged under the SM gauge group: supersymmetric models, models with
extra Higgs bosons, etc. Exhaustive studies of SM Higgs boson
production at LHC in various channels afford an opportunity of
getting for free very accurate estimates of LHC sensitivities to new
physics, and new scalar in $\gamma\gamma+jet$ channel is not the only
example.

Generally, an appropriate channel can be useful in searches for new
physics in a wider kinematical window, than the window viable for SM
Higgs boson. This occurs in models with sgoldstinos:
$\gamma\gamma$-decay mode survives for masses larger than 140~GeV,
contrary to the case of SM Higgs boson. To obtain the approximation to
the LHC sensitivity, both signal and background NLO K-factors have
been extrapolated, so this model still waits for thorough
analysis. Another source of uncertainty is related to crude estimates
of detector mass resolution for photon pairs with invariant mass
larger than 140~GeV. As we demonstrated this mass range is very
important  for new phisics. These searches will be more efficient with
better invariant mass resolution.

To summarize, this work shows that $pp\to\gamma\gamma+jet$ is very
promising in searches for both SM Higgs boson and new physics and
deserves further investigations.

{\bf Acknowledgments.} We would like to thank V.~Ilyin, N.~Krasnikov
and V.~Rubakov for useful discussions. This work was supported in part
by the RFBR02-02-17398 and GPRFNS-2184.2003.2 grants.  The work of
D.G. was also supported in part by a fellowship of the "Dynasty"
foundation (awarded by the Scientific Council of ICFPM), by the GPRF
grant MK-2788.2003.02, by a fellowship of the ``Russian Science
Support Foundation'' and by RFBR grant 04-02-17448.

\bibliographystyle{amsplain}

\end{document}